\documentclass[preprint]{sig-alternate-10pt}
\usepackage{multirow,graphicx}
\usepackage{xspace,adjustbox}
\usepackage{balance}
\usepackage{tabulary}
\usepackage{url}
\usepackage{ifthen}

\newcommand{\TITLE}{ Finding Needles in the Haystack: Harnessing Syslogs for Data
             Center Management}

\newcommand{\xxx}[1]{}
\newcommand{\todo}[1]{}
\newcommand{\sysname}{Log-Prophet\xspace}
\newcommand{\graphs}{PGs\xspace}

\newcommand{\QED}{QED\xspace}
\newcommand{\graph}{PG\xspace}

\newcommand{\syslog}{{\em syslog}\xspace}
\newcommand{\syslogs}{{\em syslogs}\xspace}

\newcommand{\tempgen}{{\em Sys\-Digest\xspace}\xspace}

\usepackage{array}
\usepackage{xspace}
\newcolumntype{L}[1]{>{\centering\let\newline\\\arraybackslash\hspace{0pt}}m{#1}}
\newcolumntype{C}[1]{>{\centering\let\newline\\\arraybackslash\hspace{0pt}}m{#1}}
\newcolumntype{R}[1]{>{\centering\let\newline\\\arraybackslash\hspace{0pt}}m{#1}}

\date{}

\author{
\begin{tabular}{cccc}
\aufnt{Chen Liang} & \aufnt{Theophilus Benson} & \aufnt{Partha Kanuparthy} & \aufnt{Yihua He} \\
\affaddr{Duke University}  &  \affaddr{Duke University} & \affaddr{Yahoo Research} & \affaddr{Yahoo} \\
\end{tabular}
}

\begin{document}


\title{\TITLE}


\maketitle


\begin{sloppypar}
\begin{abstract}


Network device syslogs are ubiquitous and abundant in modern data centers with most large data centers producing millions of messages per day. Yet, the operational information reflected in syslogs and their implications on diagnosis or management tasks are poorly understood.  Prevalent approaches to understanding syslogs focus on simple correlation and abnormality detection and are often limited to detection providing little insight towards diagnosis and resolution.

Towards improving data center operations, we propose and implement \sysname, a system that applies a toolbox of statistical techniques and domain specific models to mine detailed diagnoses. \sysname infers causal relationships between syslog lines and constructs succinct but valuable \emph{problem graphs}, summarizing root causes and their locality, including cascading problems. We validate \sysname using problem tickets and through operator interviews. To demonstrate the strength of \sysname, we perform an initial longitudinal study of a large online service provider's data center. Our study demonstrates that \sysname significantly reduces the number of alerts while highlighting interesting operational issues. 




\end{abstract}

\section{Introduction}
%

The performance of a content provider, and ultimately its revenue stream, is tied to the performance of its data center networks.  Recognizing this trend, the networking community has proposed a number of architectures, frameworks and protocols to improve data center performance~\cite{vl2:sigcomm11}, and to diagnose reliability and performance problems~\cite{Potharaju:nsdi13,dctcp:sigcomm10,vl2:sigcomm11}.  Recent efforts on diagnosing data center problems have focused on using problem tickets~\cite{Potharaju:nsdi13} and host counters~\cite{yu:sigcomm11}. Surprisingly, network device syslogs, which have proved invaluable for diagnosing problems in ISP networks~\cite{turner:imc13,turner:sigcomm10,mahimkar:sigcomm10}, have remained largely ignored in the data center space.  Most notably, works focused on using syslogs to quantify and characterize physical failures~\cite{Potharaju:socc13,Potharaju:imc13,gill:sigcomm2011,Potharaju:sigmetrics13}.  Instead of syslogs, other have used packet traces, SNMP, and ICMP to detect and diagnose problems; however, while highly powerful in detecting problems, these techniques often provide minimal aide in problem resolution.


We argue that syslogs, unlike most other readily available management data (e.g. SNMP, pings, traceroutes), are imbued with sufficient information to not only diagnose problems but also to expedite problem resolution by highlighting critical properties of the problem. Despite their potential benefits, syslogs have been ignored because they lack structure (free form text), lack homogeneity (they differ across vendors and even for a vendor, they may differ across firmware versions), and lack global causality semantics that are necessary for diagnosis (information identifying causal relationships between a haystack of messages).

Current approaches to log analysis are inappropriate for syslogs for the following reasons. First, many require access to source code~\cite{xu:detecting,xu:lprof}. Unfortunately, network devices run closed source, proprietary OSes.~\footnote{We note that even SDN capable devices come as closed source entities.} Second, blackbox techniques~\cite{oliner:dsn10,lakhina:PCA} employ simple correlation limiting their applicability and reducing accuracy.


In this work, we take the first step towards building a general log analysis framework that has two goals: (1) localize problems in the network (including problems that cascade between devices and protocol stack layers), and (2) diagnose root causes of problems (i.e., sequence of cascading events that caused the problem). Our system, \sysname,  aids operators by generating a structured representation of each problem: a causal graph of the cascading set of events (in unstructured syslogs). \sysname mines causality in syslogs to generate this output, also called \emph{Problem Graphs} (PGs).  Our work builds on two insights. First, in data centers there is an abundance of logs that naturally lend themselves to statistical techniques.  Second, network devices run networking protocols which dictate how problems propage through the network: models of these protocols can be used to guide our statistical techniques, significantly improving problem troubleshooting.


Our goal is to extract from a stream of syslog messages, a set of accurate and precise PGs that enables network operators to more readily answer, diagnosis, and resolve questions.  In this paper we take a first step towards this goal, by presenting: a prototype of \sysname; running the prototype on seven months worth of production syslog data from a large-scale data center network; and presenting a longitudinal study of the \graphs discovered in the data center.



Our key contributions are: 
\begin{itemize}
	\item \textbf{\sysname:} The design of a log analysis framework for data center networks that leverages domain knowledge and statistical technique to build problem graphs (\graphs) (Section~\ref{s:method}).
	\item \textbf{Prototype System:} A prototype implementation of \sysname allowing us to validate our design and explore its benefits. (Section~\ref{s:validation})
		
	\item \textbf{Study of \graphs in a large Data center:} An initial analysis of the \graphs generated by the data center of a large online service problem over a seven month period. We observe several interesting patterns about the \graphs generated by devices in different layers of the data center (Section~\ref{s:analysis}). 

\end{itemize}

\section{Related Work}



\textbf{Log (and Syslog) Analysis.} Previous studies of Syslog have focused on either understanding failures~\cite{Potharaju:socc13,Potharaju:imc13,gill:sigcomm2011,Potharaju:sigmetrics13,turner:imc13,turner:sigcomm10}, using syslogs as input to a machine learning algorithm for diagnosis~\cite{mahimkar:sigcomm10}, or developing techniques to extract structure from the syslogs~\cite{yamanishi:kdd05,qiu:imc10}.  Our current study builds on existing approaches~\cite{yamanishi:kdd05,qiu:imc10} and leverages their techniques to structure the data enabling us to perform a wide range of analysis.  Our analysis enriches existing work~\cite{Potharaju:socc13,Potharaju:imc13,gill:sigcomm2011,Potharaju:sigmetrics13,turner:imc13,turner:sigcomm10} that focus on link and device failures by exploring the implications of causality between syslogs.  

 Orthogonal attempts to mine general system logs for diagnosis data often rely on access to source code~\cite{xu:detecting,xu:lprof}. Unfortunately, most data center switches are closed source and thus we are unable to apply these techniques.

\textbf{Statistical-based Problem Diagnosis} Our approaches builds on correlation and causality, similarly other approaches have leveraged statical analysis to learn causal relationships or dependency graph from network data~\cite{krishnan:QED, lakhina:PCA, mahimkar:NICE, tariq:WISE, kompella:SCOPE, kandula:expose, wang:learning,oliner:dsn10}. We explore a different set of data sources with different features, which requires different models; for example, domain models to introduce structure and domain models to improve speed and accuracy of statistical techniques.


\textbf{Distributed Tracing} Distributed tracing~\cite{aguilera:distributed_black_box, attariyan:root_cause_diagnosis, fonseca:xtrace, reynolds2006pip} has long been used in distributed systems as a way of building program graphs and using them to detect problems and their root causes.   
Unlike the RPC calls in distributed systems that include either accurate time, unique IDs that allow for tracing across devices, causality, or some combination, syslogs lack this information. Specifically, there is no direct way to understand propagation of events across devices.  To overcome this, we apply statistical techniques, guided by statistical techniques, to infer causality within and across devices.

\section{Data and Background}
\label{s:data}
In this section, we present an overview of our datasets and the data center network's operational patterns.

\subsection{Data}
We collected seven months worth of \syslog data from one of Yahoo's data centers. At a high-level, the data center network is a fat-tree topology consisting of three device layers (traversing bottom-up): Top of Rack (ToR) switches, two aggregation layers and a core layer. In addition, the data center includes middleboxes serving specialized functionality. We note that this general data center structure aligns with the set of data centers studied in the prior work~\cite{benson:imc10,kandula:imc09,Potharaju:socc13,Potharaju:imc13,gill:sigcomm2011,Potharaju:sigmetrics13}.

In Table~\ref{t:data}, we describe the properties of these devices, focusing on their relative quantity, and summarizing the type and amount of data collected from each type of device.  Unsurprisingly, we observe that the data center is dominated by edge or ToR switches. Interestingly, we observe a large amount of device heterogeneity: essentially there is heterogeneity at each layer of the data center with each layer consisting of devices from multiple vendors.


\noindent\textbf{Network \syslog messages (\syslogs).} \syslogs are debugging messages about events and are generated by network devices to aid network operators in diagnosing network anomalies and in verifying the impact of maintenance operations. They capture a variety of network events ranging from failed user logins and problems with routing protocols to interface failures and CPU overheating.  These messages are free-form text -- essentially raw and unstructured. 


The network devices are configured to push \syslogs to one of several collection servers.  We collect daily snapshots from these servers and aggregated these logs to develop our corpus. 
To analyze the data, we process syslogs to impose structure and enable diagnosis (we discuss the processing steps in Section~\ref{s:analysis}).

\begin{table*}
	\begin{center}
		\begin{scriptsize}
\begin{tabular}{c|c|c|c|c|c|c}
	 {\bf Device } & {\bf Devices}& {\bf Description of} & {\bf \% of data center} & {\bf \# of Vendors} &  {\bf \% of devices}  & {\bf \% of devices} \\
	{\bf Type} & &{\bf devices} & {\bf device types}  &   & {\bf with config} & {\bf with \syslog} \\ \hline


	 TOR & $TOR_A, TOR_B, TOR_C$ & Top of Rack switches & 93\% & $>$ 1& 100\%  & 25.5\%  \\ 
	MB & $MB_A, MB_B$ &Middleboxes & 2\%& $>$ 1 &  100\% & 27.1\%  \\  
	AGG & $AGG_A, AGG_B$ & Aggregation switches & 4\% & $>$ 1&    100\% & 16.1\%  \\   
	 CORE & $CORE$ &Core routers & 0.07\% & $>$ 1&  100\% & 100 \% \\ 
\end{tabular}
\caption{Overview of Dataset.}
\label{t:data}
\end{scriptsize}
\end{center}
\end{table*}


\noindent\textbf{Configuration.} We collect nightly snapshots of device configurations, both static and device runtime outputs. We use a configuration management system that automatically captures these nightly snapshots.  These configurations are instrumental in determining if observations within the \syslog are due to transient changes or permanent changes. Moreover, the configuration allows us to reconstruct the topology and determine device and protocol adjacencies.

\subsection{Domain-Specific Models}\label{domain_model}
\sysname builds on domain models to understand the cascading of events within and across devices. These models are one-time inputs and do not require operator feedback.

\noindent\textbf{Protocol adjacencies.} Network devices are configured to run a number of protocols such as BGP, OSPF, and MPLS.  For each protocol, the device is configured to peer with other devices that are not physically adjacent.  When a protocol-specific problem occurs, the event may propagate to a device's logical peers. When an OSPF problem occurs on a device (e.g., OSPF process failure), the problem will only (initially) propagate to peers of this device (e.g. route timeouts on peers).  Motivated by this observation, we build on our prior work~\cite{benson:sigcomm11} to parse network configurations and extend it to model these adjacencies.  For each protocol in our data center, we create an {\it overlay graph} capturing these adjacencies. 

\noindent\textbf{Physical adjacencies.} Similar to protocol adjacencies, problems occurring at the physical layer (layer 1) could only propagate to physically connected devices. We leverage network topology contained in configurations to add physical adjacencies.

\noindent\textbf{Protocol delays.} In addition to leveraging domain knowledge to model spatial propagation of events, \sysname also uses domain knowledge to model temporal propagation of events across devices. Events propagate between protocol-adjacent devices in two ways; First, directly through the transfer of a message. For example, the establishment of a TCP connection for a peering session will generate syslog messages on both devices. Second, events can propagate indirectly due to the absence of an event before a timeout, which also generates a message. Examples of indirect propagation include an OSPF process failure on a device, which leads to the connection and routes on its peer to timeout several milliseconds later. In the first, messages are simultaneously generated on both devices. In the second, there is a lag between the two devices.  To capture this, we parse configuration for timeout values and use these values to bound the time taken for protocol specific events to propagate across devices.

\section{Methodology}
\label{s:method}

The goal of \sysname is to localize and find root causes of problems across the data center network. Some of these problems happen in parallel, especially at large-scale; and many of the problems cascade within and between devices and across layers in protocol stack. A key to such a diagnosis system is to output a succint but useful representation of a \emph{problem} that aids operators. We define a problem as a directed graph, whose nodes are syslog messages (semantics) and whose edges are causal relationships between them (i.e., cascades) -- we term this graph as a Problem Graph (PG). A PG describes everything that happened during a problem, starting from problems within a device and how they propagated to other device(s). As an example, a module failure (PHY) in a ToR device causes a link down problem (L2) and an OSPF problem (L3) within the device, which cause an interface (L2) issue in a connected AGG device.



In Figure~\ref{f:diagram}, we present a high-level overview of \sysname's components and the interactions between these components. \sysname consists of two classes of components: asynchronous low-volume preprocessing components which generate syslog templates and use that syslog structure to generate a causality matrix. 
The second, a set of online components ingest a live data center-wide stream of syslog messages, map them into a stream of templates, extract \emph{casual rules} from the stream, and generate \graphs from the extracted rules. The online components use relatively lightweight methods.
Decoupling the offline and online components allows our techniques to scale to large data centers and enables us to parallelize execution in \sysname.

\begin{figure}
	\begin{center}
	\includegraphics[width=3in]{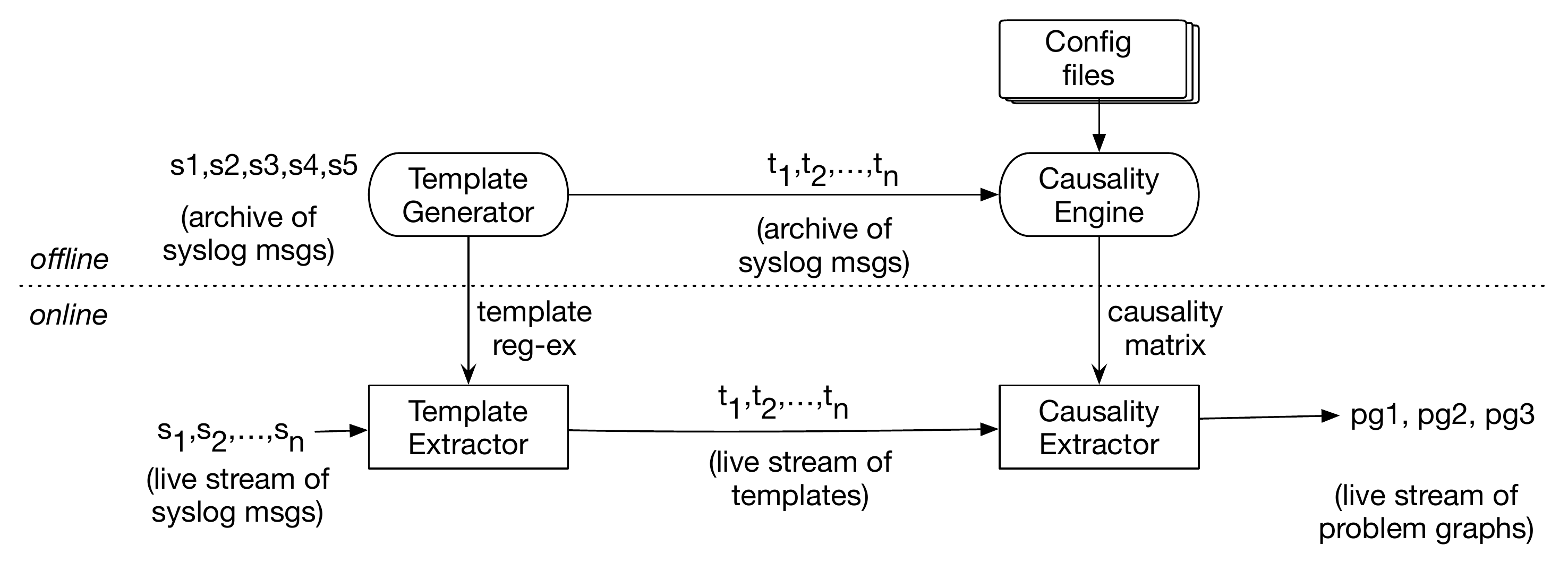}	
	\caption{Diagram of \sysname's Workflow.}
	\label{f:diagram}
	\end{center}
\end{figure}



\subsection{Extracting Structure From syslogs}
\label{s:analysis}

Several approaches have been proposed to impose structure on \syslogs and extract ``templates'' from them.  Templates are defined as short structured-format strings that succinctly describe the specific event embodied in a line of syslog as well as the entities involved in the event.  In this study, we tried several approaches and settled on the one with the least error: \tempgen~\cite{sysdigest}.  

\begin{figure}
	\begin{center}
	\includegraphics[width=3in]{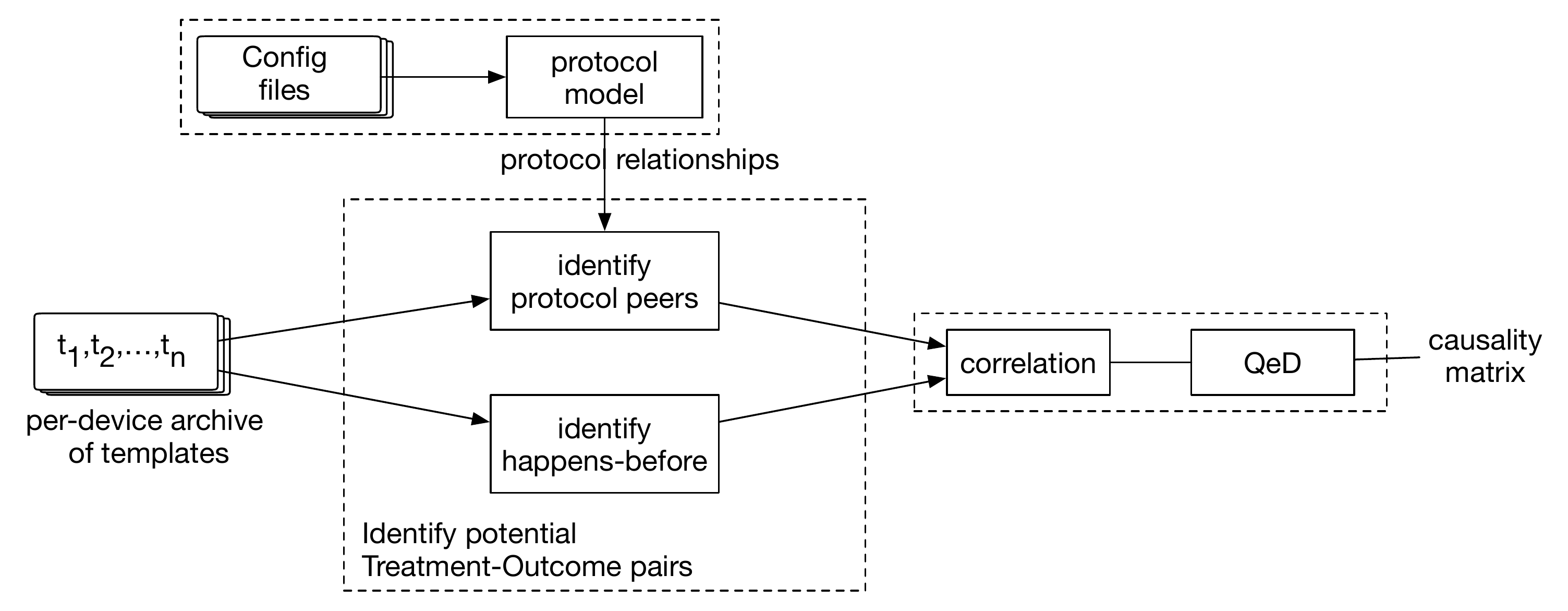}	
	\caption{Causality Engine.}
	\label{f:diagram}
	\end{center}
\end{figure}

\subsection{Causality Engine}
\sysname's causality engine, Figure~\ref{f:diagram}, is an offline component of \sysname that analyzes streams of templates to infer causality. The output of the causal engine is an  $|n \times n|$ causality matrix $M$ where $M_{ij}$ is $1$ if and only if template $t_i$ has a causal impact on $t_j$. 

In networks and distributed systems, the traditional approach for extracting causality from messages (or logs) is to implement logical timestamps in events and infer happens-before relationships. We could then use semantic information in the message (or logs) to track problems between devices. For example, Sherlock uses IP addresses in packets to determine the flow of causality where as Mystery Machine learns causality between RPCs based on happens-before relationships. These methods either require precise timing information from synchronized clocks, and/or significant instrumentation to add information such as logical clocks to device events. Since we cannot modify (closed source) network devices to add instrumentation such as logical clocks or record fine-grained timestamps, we cannot use traditional methods; further, happens-before relationships cannot be inferred from syslogs in a large-scale network (the needle in a haystack problem).



To address these challenges, our causality engine leverages the fact that data centers generate millions of events per day and are thus amenable to statistical techniques to find causality (a stronger definition than happens-before).  We apply Quasi-Experimental Design, \QED, a popular technique used by social scientists to find causal relationships in data sets~\cite{william2002experimental,oktay2010causal}. QED is often used in the absence of traditional data sources and controlled experiments such as $A|B$ testing.
\QED establishes a causal relationship between pairs of templates (from same or different devices), a treatment variable (cause), $t_i$, and an outcome variable (effect), $t_j$, by using history; and accounts for confounding factors.  For example, using \QED we are able to affirm a causal relationship between a link failure (treatment) and a routing protocol outage (outcome), conditioning for confounding factors such as device config changes, power, firmware, and operating system. Using \QED, we are able to build a statistically significant causality matrix $M$.
%


\subsection{QED}
At a high level, \QED detects causality between an independent variable X ``treatment'' and a dependent variable Y ``outcome'' by identifying a set of ``untreated'' variables $Z={z_1 ... z_n}$ that are identical to X, after accounting for confounding factors~\cite{william2002experimental}.
%
%
The key to accurately affirming causality, lies in precisely identifying all confounding factors and asserting that X is identical to elements of $Z$ --  this assertion allows us to affirm that any difference in outcomes between paired experiments on $z_i \subset Z$ and X is entirely due to X.

Thus to apply QED, the causality engine must take the following steps: (1) identify dependent variables in our data set, (2) control for a pre-defined set of confounding factors, (3) identify ``untreated'' variables for each potential causal relationship, and (4) perform a statistical test of relevance.
%
%

\textbf{Identifying dependent variables.}
To identify dependent variables, we leverage domain knowledge discussed in Section~\ref{domain_model} to identify templates across devices that may be dependent. Specifically, two templates across devices may be dependent if they have spatial and temporal adjacencies. For templates within a device, we also leverage happens-before relationships, because syslog stream from a device is strictly ordered.  Once we identify the sets of potentially dependent variables we apply nonparametric correlation to ensure statistical dependence: we correlate pairs of templates and identify two templates $t_i$ and $t_j$ as dependent if the correlation value is significant (over $alpha$). 

\textbf{Identifying confounding factors.} The confounding factors pair templates $t_i$ and $t_j$ conditioned on external factors that impact $t_i$ and $t_j$.  These confounding factors include device vendor, OS, location in data center topology, and configuration changes. Thus for each template $t_i$, we identify a confounding set $K={t_j...t_m}$ such that each element $t_j \in K$ is correlated with $t_i$ ($C_{ij} > alpha$) and they share identical confounding factors.

\textbf{Identifying untreated variables.} Given the confounding set $K$ an ``untreated'' template, $t_u$, is a randomly selected member of $K$.

\textbf{Statistical test of relevance.} Next, we compare the untreated template, $t_u$, and with the treated template, $t_i$. Using historical data, \sysname performs a hypothesis test for the null hypothesis ($H_0$) that $t_t$ has no impact on $t_o$ and calculate the $p\_value$. 
Based on the test, we affirm that $t_t$ (treatment) has a causal impact on $t_o$ (outcome).

\subsection{Constructing problem graphs (\graphs)}
\label{s:graph_build}


\begin{figure}
	\begin{center}
	\includegraphics[width=2.4in]{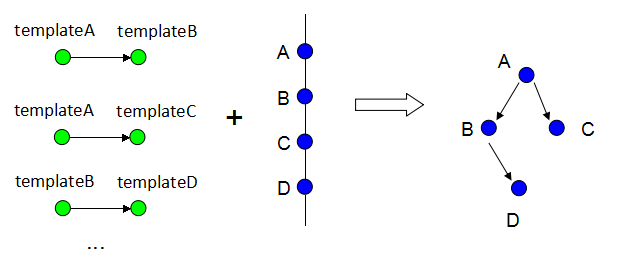}
	\caption{Causally related template pairs and a stream of templates are used to build a \graph.}
	\label{f:causal_graph}
	\end{center}
\end{figure}

Given a causallity matrix, $M$, \sysname can process a stream of templates and identify pair of templates $t_j$ and $t_i$ where $t_j$ causally impacts $t_i$ (illustrated in Figure~\ref{f:causal_graph}). We summarize the stream as a set of \graphs, encoding a set of causally related templates. 

\begin{figure}
	\begin{center}
	\includegraphics[width=1.8in]{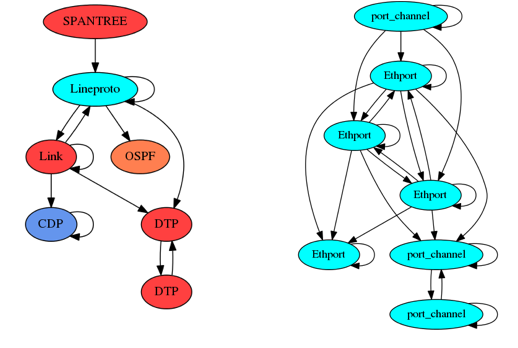}	
	\caption{Example \graphs from our data center.}
	\label{f:causal_graph_example}
	\end{center}
\end{figure}

More formally, a \graph is a graph $G = (V, E)$ with one vertex per syslog template and directed edges between syslog templates representing causality. A directed edge in a \graph represents a causal relationships between two templates (or syslog lines) with the direction indicating causality.  Figure~\ref{f:causal_graph_example} shows two examples of problems from our data center (colors encode layer in protocol stack). The left \graph shows an example of a multi-layer problem that spans AGG and ToR layers and multiple layers in the protocol stack. It encodes templates that capture a module failure template that causes a link down template, which further triggers a spanning tree protocol status change template, and at the time causing a interface status change template on a peering device. An edge from a node to itself implies that multiple, often duplicate, syslog lines are often generated by a specific event: the edge between ``LineProtocol'' and itself indicates that when LineProtocol fails, devices generate multiple syslog lines. The right \graph shows an example of a problem within ToR devices that is an Ethernet (L2) flapping issue. 

To contruct \graphs (illustrated in Figure~\ref{f:causal_graph}), \sysname processes a stream of syslog templates online, and divides the stream into time windows of templates of size $delta$.  For each window, it creates vertices for each template and adds them into graph $P$.  It adds a direct edge between $i \subset P$ and $j \subset P$ if $M_ij$ is 1.  Finally, \sysname scans $P$ for duplicate vertices $x$ and $y$ such that $template(x) == template(y)$. It merges the nodes and preserves the directed edges. If an edge exists between $y$ and $x$, then it creates a self loop.

Broadly, when used for problem determination, the key aspect of the \graph are the edges and the roots of the graph. The roots allows operators to determine (trace back) potential root cause(s) and the edges enable operators to understand the escalation/propagation of the problem.

\section{Prototype And Validation}
\label{s:validation}
We developed an initial prototype of \sysname in Python in approximately 4000 LoCs.  Our prototype builds on the statistical libraries provided by scikit-learn~\cite{scikit-learn}  for statistical analysis.  We include models for the following protocols: BGP, VLAN, and OSPF.

\textbf{Validation:}  To validate \sysname, we examined the problem ticket database for the data center and interviewed operators of our data centers. For our operator interviews, we analyzed a subset of the data: 6 weeks worth of data.  We presented the raw syslogs along with the discovered \graphs, and discussed the information contained within the \graphs as well potential problem tickets generated for these graphs.  Our initial results are promising -- the operators agreed that the \graphs succinctly captured events within a device. For events across devices, we were limited by the set of protocols we have currently modeled. We plan to extend our prototype with protocol models. We are also working on a systematic validation of \sysname output, both using live data (in collaboration with network operations), and an emulation testbed.

\section{Initial Characterization of Production \graphs}
\label{s:analysis}

In this section, we present a longitudinal analysis of syslogs and resulting \graphs for a large data center consisting of over five thousand devices. In analyzing these \graphs, we aim to understand: (1) the data center's operational dynamics from the perspective of the network devices; (2) the persistence of operations issues (as captured by \graphs); and (3) the locality of operational issues to specific types of devices or layers in the networking stack. 

\subsection{Classification of \graphs}
In applying \sysname on seven months of data, we discovered 4264 \graphs which could be easily clustered into 156 classes: through manual inspection we were able to classify these graphs into 22 high level types of \graphs.  The manual inspection allowed us to group together graphs consisting of templates from different vendors -- the clustering couldn't not do this as their graphs were included radically different templates.

We observe that applying \sysname to our data center allows us to reduce hundreds of millions of lines into 13 distinct \graphs with each \graph. We analyzed a time series of the number of syslogs generated and compare them against the number of instances of \graphs extracted (not shown due to space constraints). From this we observe that while there are only 13 unique \graphs in our data center: there are in fact thousands of events occurring on a daily basis.  From an operational standpoint, this is a more tractable number than the hundreds of millions of syslogs. Furthermore, while there are thousands of instances \graphs they only map to thirteen unique \graphs thus problem determination and resolution will involve little overhead.

\begin{figure}[h]
  \begin{center}
    \includegraphics[width=3.4in]{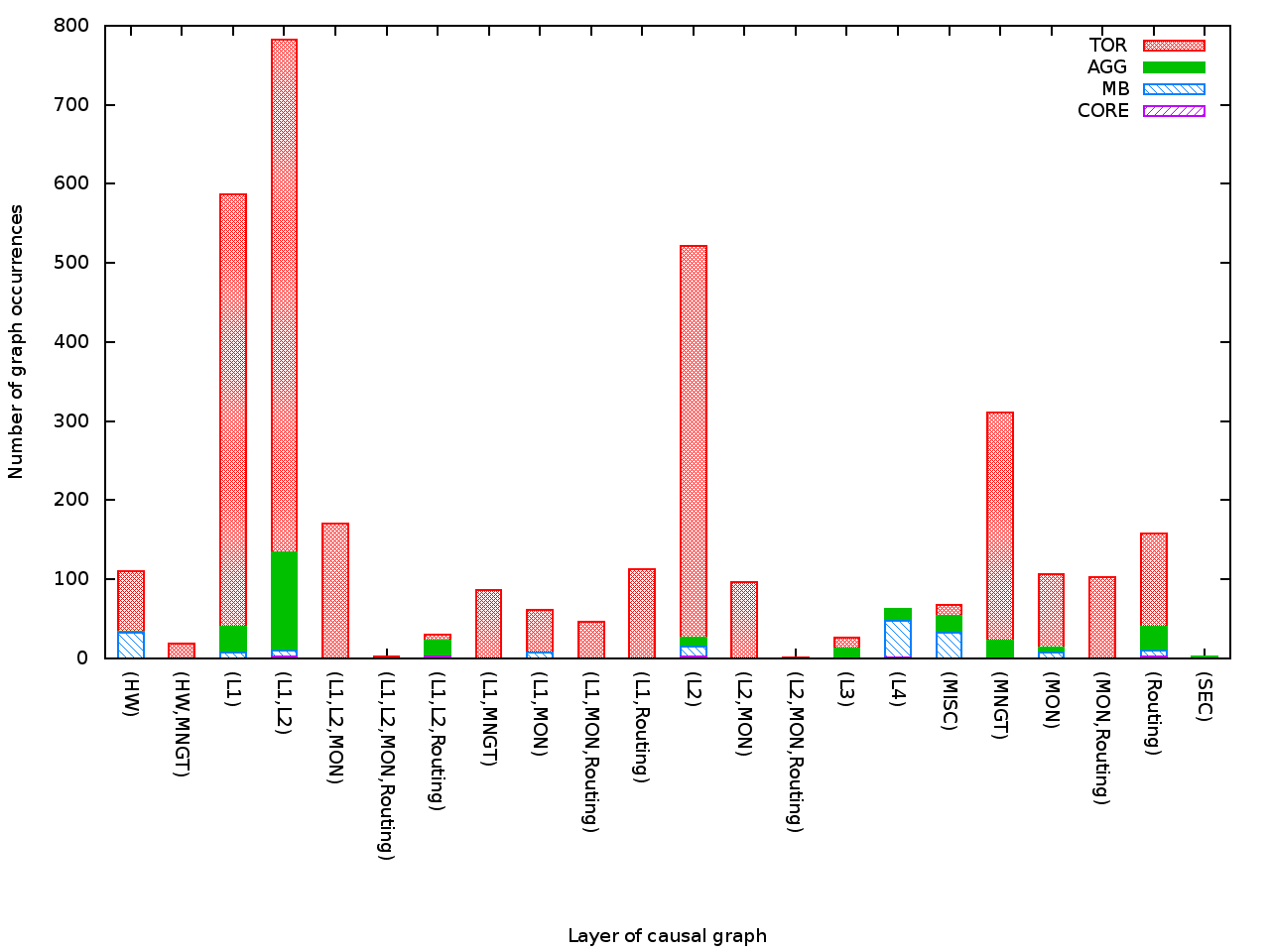}
    \caption{Number of \graphs for each layer.}
    \label{f:gsig_cnts_layer}
  \end{center}
\end{figure}

\subsection{Footprint of \graphs}
Next in Figure~\ref{f:gsig_cnts_layer}, we analyze the breakdown of \graphs across the different layers in the data center. There are several interesting observations about the distribution of graphs across data centers layers.

 \textbf{ToR layer dominates:} Over 93\% of the \graphs occur at the ToR devices. This pattern is a direct manifestation of two trends: first, over 90\% of the devices in the data center are ToR devices (See Table~\ref{t:data}); second, ToR devices are often cheap commodity devices with lower quality parts and are thus prone to more failures~\cite{Potharaju:imc13,gill:sigcomm2011}.

\textbf{L3-L4 \graphs dominate MBs:} Unlike the ToR and the AGG layers, the middlebox layer are dominated by a combination of L4 \graphs (virtual IP migration), routing \graphs, and HW \graphs. In the the different types of MB only generate 2-3 graphs: for example, the load balancer generates \graphs related to: virtual IP migration (or failures); memory issues; and interface failures creating IP unreachability.

\textbf{\graphs across devices:} A number of interesting \graphs display problems that cross devices boundaries and even tier boundaries.  For example, we find that graphs related to interface failures in a ToR often percolate up to affect the OSPF routing protocol in adjacent AGG devices (illustrated in Figure~\ref{f:causal_graph_example}).

	\begin{figure}[h]
	  \begin{center}
	    \includegraphics[width=2.8in]{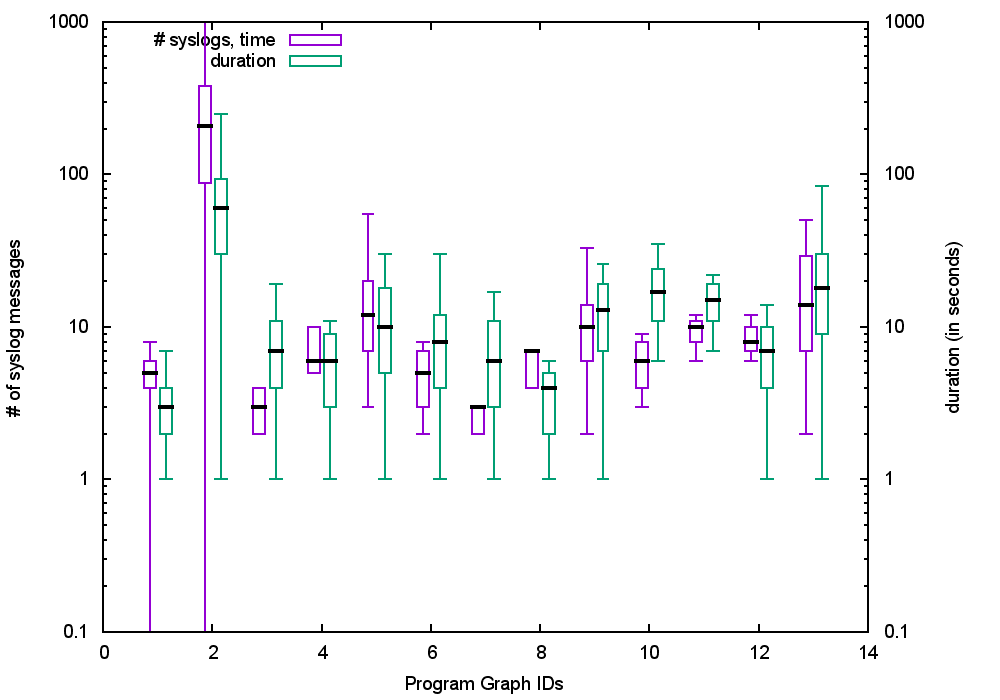}
	    \caption{Number of \graphs for each layer.}
	    \label{f:candlesticks}
	  \end{center}
	\end{figure}
Lastly in Figure~\ref{f:candlesticks}, we analyze the size of a representative subset of the \graphs in terms of duration (in seconds) and length (number of messages covered). Due to space we limit the figure to 14 representative \graphs. Naturally, there exists a diversity in the sizes of the different \graph.  More interestingly, we observe that counter intuitively, there is no direct correlation between duration and syslog lines capture: this indicates that simply looking at number of syslogs is insufficient. 

\textbf{Takeaways:} Although our results are preliminary, the provide us with an overview of the problems with modern data centers. We are currently working with operators to explore methods for developing tools to automatically suppress superflous \graphs; ranks \graphs based importance and scope; and to use \graphs to suggestion potential resolutions.

\balance\section{Conclusion}
\label{s:conclusion}
There exists a dichotomy within the community. On one hand, a large number of studies have characterized data center failures and performance issues using a specific set of syslog messages (i.e. interface failures). On the other hand, extensive work within carrier and ISP networks have displayed the benefits of using a larger set of syslog messages. In this work, we take the first step towards reconciling this dichotomy by developing \sysname, a system for analyzing syslog messages and extracting problem graphs that summarize causality within the data center. We have validated \sysname with network operators and applied to an initial data center.  Our initial analysis are promising and they show that there is a strong benefit in employing frameworks such as \sysname.
\balance\label{lastpage}
\newpage
\end{sloppypar}



\small
\setlength{\parskip}{-1pt}
\setlength{\itemsep}{-1pt}
\bibliography{references}
\bibliographystyle{abbrv}

\end{document}